\documentclass[12pt,a4paper]{article}

\usepackage{epsfig}
\usepackage{subfigure}
\usepackage{graphicx}
\usepackage{amsmath}

\begin{document}
\begin{titlepage}
\begin{flushright}
CERN-TH-2017-008
\end{flushright}
\begin{center}

{\large \bf {A Simple Method to detect spontaneous CP Violation in multi-Higgs models}}

\vskip 1cm

O. M. Ogreid,$^{a,}$\footnote{E-mail: omo@hvl.no}
P. Osland$^{b,}$\footnote{E-mail: Per.Osland@uib.no} and 
M. N. Rebelo$^{c,d, }$\footnote{E-mail: rebelo@tecnico.ulisboa.pt}

\vspace{1.0cm}

$^{a}$Western Norway University of Applied Sciences, \\
Postboks 7030, N-5020 Bergen, Norway, \\
$^{b}$Department of Physics and Technology, University of Bergen, \\
Postboks 7803, N-5020  Bergen, Norway,\\
$^{c}$Centro de F\'isica Te\'orica de Part\'iculas -- CFTP and Dept de F\' \i sica\\
Instituto Superior T\'ecnico -- IST, Universidade de Lisboa, Av. Rovisco Pais, \\
P-1049-001 Lisboa, Portugal \\
$^{d}$ Theoretical Physics Department, CERN, CH-1211, Geneva 23, Switzerland \\

\end{center}

\vskip 3cm

\begin{abstract}
For models with several Higgs doublets
we present an alternative method to the one proposed by Branco, Gerard and 
Grimus, in 1984, to check whether or not CP is 
spontaneously violated in the Higgs potential. The previous 
method is powerful and rigorous. It requires the identification of a matrix $U$ 
corresponding to a symmetry of the Lagrangian and verifying a simple relation 
involving the vacuum  expectation values. The nonexistence of such a matrix
signals spontaneous CP violation. This approach may be far from trivial as complexity grows with the number of Higgs doublets. In such cases it may turn out to be easier to analyse the potential by going to the 
so-called Higgs basis.  The transformation to the Higgs basis is straightforward 
once the vacuum expectation values
are known. The method proposed in this work is also powerful and
rigorous and can be particularly useful to analyse models with more than two Higgs
doublets and with continuous symmetries.
\end{abstract}

\end{titlepage}

\noindent
{\it Spontaneous CP Violation} \\
Models with more than one Higgs doublet allow for the possibility
of having spontaneous CP violation. The idea of spontaneous T (hence CP)  violation was 
first proposed by T. D. Lee  \cite{Lee:1973iz} in the context of two-Higgs-doublet 
models. CP can only be spontaneously violated if the Lagrangian is invariant 
under CP and if at the same time there is no transformation that can be identified with a 
CP transformation leaving both the Lagrangian and the vacuum invariant. 
In the Standard Model  there is only one Higgs doublet and the scalar potential 
necessarily conserves CP. 

Under a CP transformation a single Higgs doublet, $\Phi$, transforms into 
its complex conjugate.
In the presence of more than one doublet the most general CP transformation 
\cite{Branco:1999fs}
allows for mixing of the scalar doublets under an arbitrary unitary matrix, $U$:
\begin{equation}
\Phi_i \stackrel{\mbox{CP}}{\longrightarrow} U_{ij} \Phi^\ast_j
\label{abc}
\end{equation}
This transformation combines the CP transformation of each Higgs doublet with 
a Higgs basis transformation.  Higgs basis transformations do not change the 
physical content of the model. If the potential is invariant under such a 
transformation there is explicit CP conservation. At this stage $U$ applied
to the $\Phi_j$ fields is not required to be a symmetry of the Lagrangian.
It is trivial to see that when all the coefficients 
of the potential are real the above condition is verified by a matrix $U$ 
equal to the identity and CP is not violated explicitly. 

In multi-Higgs
models it may not be trivial to check whether CP is violated explicitly or not
in the scalar sector due to the freedom one has to make Higgs basis
transformations. These transformations change the quadratic and quartic 
couplings and in particular couplings that are complex in one basis
may become real in another, and vice versa. This fact has motivated the study of 
conditions for CP invariance at the Lagrangian level expressed in terms of CP-odd Higgs-basis invariants
\cite{Branco:2005em, Gunion:2005ja}.

Once it is established that the potential does not violate CP explicitly the 
question remains of whether or not there is spontaneous CP violation.  
It has been shown \cite{Branco:1983tn} that the vacuum is CP invariant if the 
following relation is verified with a matrix $U$ corresponding to a symmetry
of the Lagrangian:
\begin{equation}
 U_{ij} \langle 0| \Phi_j |0\rangle^\ast = \langle 0| \Phi_i |0\rangle
\label{geo}
\end{equation}
This is a very powerful relation. 
It is stated in Ref.~\cite{Branco:1983tn} that: given a particular set of 
vacuum expectation values (vevs) the 
simplest way of proving  that they do not break  CP 
is to construct a unitary matrix $U$ which satisfies Eq. (\ref{geo})
and which corresponds at the same time to a symmetry of the Lagrangian. 
This prescription is rigorous but in some cases the construction of this
matrix may not be obvious. \\

If such a difficulty arises we propose a simple test which proves useful
in identifying CP-conserving cases.
Once the set of vevs is determined, we go to the so-called 
``Higgs basis" defined as a basis where only one of the Higgs doublets 
acquires a vev  different from  zero and chosen to be real. It is 
straightforward to build a transformation that takes the fields to such a special 
basis \cite{Donoghue:1978cj, Georgi:1978ri},
by means of the product of an orthogonal matrix by a diagonal matrix with phases 
equal to those of the
original vevs but with opposite sign. It should be pointed out that
for more than two Higgs doublets such a basis is defined up to a unitary
transformation of the new $(n-1)$ doublets with zero vev.
If the coefficients of the scalar potential 
in such a basis can be made real by means of the rephasing freedom that is 
still left for the doublets with zero vevs, we may conclude
that CP is not spontaneously broken.\footnote{We thank H. Haber for pointing 
out to us that in some cases this unitary $(n-1) \times (n-1)$ matrix may play a role
in making the potential real.} 
Obviously, in this case, once in such a Higgs basis, we may define a CP transformation 
given by Eq.~(\ref{abc}) that verifies the relation
given by Eq.~(\ref{geo}) by simply choosing the matrix $U$ to be diagonal.
If on the contrary it proves impossible to make the scalar potential real 
in such a Higgs basis we must make sure
that we are not in one of the special cases of CP conservation
with irremovable complex coefficients  \cite{Ivanov:2015mwl}
before concluding that CP is violated.
On the other hand, this procedure, complemented with the use of CP-odd invariant 
conditions \cite{Branco:2005em, Gunion:2005ja, Lavoura:1994fv,Botella:1994cs}
may also prove useful to confirm the existence of  CP violation since in this case
it must be possible to find CP-odd invariants that are non-zero.\footnote{The technique involved in deriving CP-odd invariant conditions was introduced for the first time in Ref.~\cite{Bernabeu:1986fc} and subsequently applied in many different contexts.} \\

\noindent
{\it  Special cases in the framework of three-Higgs-doublet models 
with $S_3$ symmetry.} \\
CP conserving scalar potentials with irremovable phases are very special and rare.
Imposing explicit CP conservation in the $S_3$-symmetric three-Higgs-doublet model
by taking all parameters of the potential real does not lead to loss of 
generality \cite{Ivanov:2015mwl} and was adopted in 
Ref.~\cite{Emmanuel-Costa:2016vej} where a detailed study of the possible
vacua of the $S_3$-symmetric three-Higgs-doublet potential is performed
with emphasis on the cases in which the CP symmetry can be spontaneously broken. 
Different vacuum solutions correspond to different regions of parameter space which are
identified in Ref.~\cite{Emmanuel-Costa:2016vej}.\\

First, we illustrate some of the features of the Higgs basis by analysing a 
special complex solution for the vevs of the scalar potential written in terms 
of the $S_3$ defining representation, i.e., three Higgs doublets such that
the potential is invariant under any permutation of these fields. 
This representation is known to be reducible. The scalar 
potential ($V=V_2+V_4$) acquires the following form  \cite{Derman:1978rx}: 
\begin{subequations}
\label{Eq:pot-original}
\begin{align}
V_2&=-\lambda\sum_{i}\phi_i^\dagger\phi_i +
\frac{1}{2}\gamma\sum_{i<j}[\phi_i^\dagger\phi_j+ \mbox{h.c.} ],\\
V_4&=A\sum_{i}(\phi_i^\dagger\phi_i)^2
+\sum_{i<j}\{C(\phi_i^\dagger\phi_i)(\phi_j^\dagger\phi_j)
+\overline C (\phi_i^\dagger\phi_j)(\phi_j^\dagger\phi_i) 
+\frac{1}{2} D[(\phi_i^\dagger\phi_j)^2+\mbox{h.c.}] \} \nonumber \\
&+\frac{1}{2} E_1\sum_{i\neq j}[(\phi_i^\dagger\phi_i)(\phi_i^\dagger\phi_j)
+ \mbox{h.c.} ]
+\sum_{i\neq j\neq k\neq i,j<k}
\{\frac{1}{2} E_2[(\phi_i^\dagger\phi_j)(\phi_k^\dagger\phi_i)+ \mbox{h.c.} ]
\nonumber \\
&+\frac{1}{2} E_3[(\phi_i^\dagger\phi_i)(\phi_k^\dagger\phi_j)+ \mbox{h.c.} ]
+\frac{1}{2} E_4[(\phi_i^\dagger\phi_j)(\phi_i^\dagger\phi_k)+ \mbox{h.c.} ] \}.
\end{align}
\label{29ab}
\end{subequations}
It was pointed out long ago \cite{Derman:1979nf} that a possible complex vacuum solution is 
given by $(x, xe^{\pm\frac{2\pi i}{3}}, xe^{\mp\frac{2\pi i}{3}})$.
This solution was discussed in \cite{Branco:1983tn}. It has the remarkable feature of
corresponding to a set of vevs with calculable non-trivial phases assuming geometrical values, 
i.e., fixed values that are not expressed as functions of the parameters of the potential, and 
which are  entirely determined by the symmetry of the scalar potential. These phases cannot 
be removed by a simple rephasing of the Higgs fields, while at the same time
keeping the coefficients of the Higgs potential real. However
they do not lead to spontaneous CP violation \cite{Branco:1983tn}
since there is a matrix $U$ satisfying the constraint of Eq.~(\ref{geo}), namely:
\begin{equation}
U = \left( \begin{array}{ccc}
1 & 0 & 0 \\
0 & 0 & 1 \\
0 & 1 & 0 
\end{array} \right),
\end{equation}
which is at the same time a symmetry of the potential. It looks, in fact, as if we are 
in the presence of irremovable CP conserving phases. However, there is a nontrivial unitary 
transformation giving rise to real vevs together with a real potential:
\begin{equation}
\left( \begin{array}{c}
\phi_1^\prime \\
\phi_2^\prime \\
\phi_3^\prime \\
\end{array}  \right) = \left( \begin{array}{ccc} 
\frac{1}{\sqrt{3}}( 1 & 1 & 1) \\
\frac{1}{\sqrt{2}}( 1 & -1 & 0) \\ 
\frac{1}{\sqrt{6}}( 1 & 1 & -2) \\
\end{array} \right) \left( \begin{array}{ccc} 
1 & 0 & 0 \\
0 & e^{\mp i  \frac{2 \pi}{3}} & 0 \\
0 & 0 & e^{ \pm i  \frac{2 \pi}{3}} 
\end{array} \right) 
 \left( \begin{array}{c}
\phi_1 \\
\phi_2 \\
\phi_3 \\
\end{array}  \right)
\label{basis0}
\end{equation}
This transformation takes us to the Higgs basis with only one nonzero real vev. Notice
that this unitary transformation is also given by geometrical angles, in the sense 
defined above. The fields $\phi_2^\prime$ and $\phi_3^\prime$ acquire zero vevs and can 
be rephased to remove unwanted phases from the potential so that the potential is real.
Obviously, the transformation to the Higgs basis always
consists of the product of an orthogonal matrix with a diagonal matrix 
with phases. These phases are the complex conjugates of the phases of the
vevs of each doublet in the initial basis. The orthogonal matrix
has one row given by $(1/N) (|v_1|, |v_2|,. . ., |v_n|)$, where $n$ is 
the number of doublets and $1/N$ a normalisation factor. Finally, there is
still freedom to rephase the new doublets with zero vev.  \\

The scalar potential written in terms of the $S_3$ irreducible representations singlet
and doublet fields, respectively  ($h_S$) and ($h_1, h_2$), can be written 
\cite{Kubo:2004ps,Teshima:2012cg,Das:2014fea}:
\begin{subequations}
\begin{align}
V_2&=\mu_0^2 h_S^\dagger h_S +\mu_1^2(h_1^\dagger h_1 + h_2^\dagger h_2), \\
V_4&=
\lambda_1(h_1^\dagger h_1 + h_2^\dagger h_2)^2 
+\lambda_2(h_1^\dagger h_2 - h_2^\dagger h_1)^2
+\lambda_3[(h_1^\dagger h_1 - h_2^\dagger h_2)^2+(h_1^\dagger h_2 + h_2^\dagger h_1)^2]
\nonumber \\
&+ \lambda_4[(h_S^\dagger h_1)(h_1^\dagger h_2+h_2^\dagger h_1)
+(h_S^\dagger h_2)(h_1^\dagger h_1-h_2^\dagger h_2)+ \hbox {h.c.}] 
+\lambda_5(h_S^\dagger h_S)(h_1^\dagger h_1 + h_2^\dagger h_2) \nonumber \\
&+\lambda_6[(h_S^\dagger h_1)(h_1^\dagger h_S)+(h_S^\dagger h_2)(h_2^\dagger h_S)] 
+\lambda_7[(h_S^\dagger h_1)(h_S^\dagger h_1) + (h_S^\dagger h_2)(h_S^\dagger h_2) 
+\hbox {h.c.}]
\nonumber \\
&+\lambda_8(h_S^\dagger h_S)^2.
\label{Eq:V-DasDey-quartic}
\end{align}
\end{subequations}
The irreducible representations can be related to the reducible-triplet fields by:
\begin{equation}
\left( \begin{array}{c}
h_S \\
h_1\\
h_2\\
\end{array}  \right) = \left( \begin{array}{ccc} 
\frac{1}{\sqrt{3}}( 1 & 1 & 1) \\
\frac{1}{\sqrt{2}}( 1 & -1 & 0) \\ 
\frac{1}{\sqrt{6}}( 1 & 1 & -2) \\
\end{array} \right) 
 \left( \begin{array}{c}
\phi_1 \\
\phi_2 \\
\phi_3 \\
\end{array}  \right),
\end{equation}
and all coefficients of the potential remain real.
The translation of the coefficients of the potential from one 
framework to the other is given explicitly in Ref.~\cite{Emmanuel-Costa:2016vej}.
Notice the similarity of this transformation with the one given in Eq.~(\ref{basis0}). 
In this basis the vevs of $(h_1, h_2, h_S)$ corresponding to the 
previous solution $(x, xe^{\pm\frac{2\pi i}{3}}, xe^{\mp\frac{2\pi i}{3}})$
are of the form \cite{Emmanuel-Costa:2016vej} 
$(w_1, w_2, w_S) = (\hat w_1,\pm i\hat w_1,0)$ 
with $\hat w_1$ real and positive by convention. A phase 
redefinition of $(h_1, h_2, h_S)$ 
through  $\mbox{diag}( 1, i, 1)$ gives rise to real vevs and as a result the term
in   $\lambda_4$ splits into different terms with coefficients 
$i\lambda_4$ or $-i\lambda_4$, i.e., no longer real. All these coefficients 
can be made real 
by multiplying $h_S$ by $i$, note that $h_S$ has zero vev. At this stage 
both the vevs and coefficients of the potential are real. To go to the Higgs 
basis one still needs a rotation that mixes $h_1$ and $h_2$. This rotation
is real and therefore it does not introduce any new phase. \\

Now, we revisit some of the solutions with complex vacua and $\lambda_4 = 0$
discussed in Ref.~\cite{Emmanuel-Costa:2016vej}. As pointed out there,
in the framework of three-Higgs-doublet models with $S_3$ symmetry 
spontaneous CP violation cannot occur if $\lambda_4 = 0$. Furthermore, for
$\lambda_4 = 0$ the potential acquires an additional SO(2) symmetry.

A particularly interesting vacuum is the one identified as case C-III-c, which 
in terms of the irreducible representations is of the form   
$(\hat {w_1} e^{i\sigma},\hat{w_2},0)$.  This complex vacuum 
requires that
three constraints among the coefficients of the potential be verified, one
of them being $\lambda_4 = 0$. At first sight it looks as if it violates
CP spontaneously, due to the fact that the moduli of $w_1$ and of $w_2$ 
are different. Clearly, there is no obvious simple form for the matrix $U$ 
satisfying the constraint of Eq.~(\ref{geo}). Therefore, the easiest and most 
straightforward way of checking for CP conservation is to look at the potential
in the Higgs basis, which can be reached via the simple transformation:
\begin{equation}
\left( \begin{array}{c}
h_1^\prime \\
h_2^\prime \\
h_S^\prime \\
\end{array}  \right) = \frac{1}{v} \left( \begin{array}{ccc} 
\hat {w_1}  & \hat {w_2} & 0 \\
 \hat {w_2} & - \hat {w_1}& 0 \\ 
 0 & 0 & v \\
\end{array} \right) \left( \begin{array}{ccc} 
e^{-i\sigma} & 0 & 0 \\
0 & 1 & 0 \\
0 & 0 & 1 
\end{array} \right) 
 \left( \begin{array}{c}
h_1 \\
h_2 \\
h_S \\
\end{array}  \right)
\label{basis2}
\end{equation}
with $v^2= (\hat {w_1}^2 + \hat {w_1}^2)$ and confirming by inspection that 
the coefficients of the potential remain real, while now
all vevs are real. Notice that in the Higgs basis 
there is freedom to rephase $h_2^\prime$ and $h_S^\prime$.

The construction of the matrix $U$ satisfying the constraint of Eq.~(\ref{geo})
was presented in Ref.~\cite{Emmanuel-Costa:2016vej} and makes use of the
additional $SO(2)$ symmetry resulting from having  $\lambda_4 = 0$:
\begin{equation}
U = 
e^{i(\delta_1+\delta_2)}
\begin{pmatrix}
\cos\theta &\sin\theta & 0\\
-\sin\theta & \cos\theta & 0 \\
0 & 0 & 1
\end{pmatrix}
\begin{pmatrix}
0 & 1 & 0 \\
1 & 0 & 0 \\
0 & 0 & 1
\end{pmatrix}
\begin{pmatrix}
\cos\theta &-\sin\theta & 0 \\
\sin\theta & \cos\theta & 0 \\
0 & 0 & 1
\end{pmatrix}
\label{u2}
\end{equation}
where the angle $\theta$ is such that the matrix on the right-hand side 
of Eq.~(\ref{u2}) rotates $(\hat w_1 e^{i \sigma}, \hat w_2, 0)$  into vevs of 
the form $(ae^{i\delta_1},ae^{i\delta_2},0)$, where the two nonzero entries
have the same modulus. This is possible due to the additional $SO(2)$
symmetry and requires 
$\tan 2 \theta = (\hat w_1^2-\hat w_2^2) /  (2\hat w_1\hat w_2 \cos\sigma)$.
An overall rotation by the phase factor $\exp[-i(\delta_1+\delta_2)/2]$ 
leads then to vevs of the form $(ae^{i\delta},ae^{-i\delta},0)$. The matrix $U$ 
also makes use of the symmetry under the interchange 
$h_1^\prime \leftrightarrow h_2^\prime$, as can be seen from 
the matrix in the middle. 
Equations (\ref{basis2}) and (\ref{u2}) have in common the fact 
that both rotations depend on the vevs of the Higgs doublets. Once the vevs
are known a rotation to the Higgs basis can be easily determined. However
building the matrix $U$ requires insight and therefore there is the 
possibility of missing it in cases where in fact CP is conserved
since there is no well-defined prescription to build it. On the other hand, 
once CP conservation is established it follows that
the matrix $U$ of Eq.~(\ref{geo}), corresponding to a symmetry of the 
Lagrangian, must exist. \\

\noindent
As mentioned above, in the case of $\lambda_4=0$ the 
$S_3$-symmetric potential acquires an 
additional $SO(2)$ symmetry. Spontaneous breaking of this symmetry 
leads to a massless scalar field, which is ruled out by experiment.
This problem can be avoided by adding soft breaking terms to the potential.
Soft breaking terms  of the form ($h_S^\dagger h_i + h.c.$) 
are only consistent, once the minimisation conditions are imposed, 
if their coefficients are proportional to $\lambda_4$,
therefore, in this case, we are only left with the possibility of adding
terms of the form $\mu_2^2(h_1^\dagger h_1 - h_2^\dagger h_2)+
\frac{1}{2}\nu^2(h_2^\dagger h_1 + h_1^\dagger h_2)$. It has been checked 
that the potential with these additional terms still allows for vevs of the
form C-III-c and that the transformation to the Higgs basis 
together with rephasing of the fields with zero vevs can lead to a new 
potential with only real coefficients. \\

\noindent
The only other complex vacuum with  $\lambda_4 = 0$ having non trivial phases,
i.e., phases differing from $\pm i$ is case C-IV-e which has the form
($\sqrt{-\frac{\sin 2\sigma_2}{\sin 2\sigma_1}}\ \hat w_2e^{i\sigma_1}, 
\hat w_2e^{i\sigma_2},\hat w_S$).
In this case
there is no zero vev and all vevs have different moduli. The construction of 
a matrix $U$ satisfying the constraint of Eq.~(\ref{geo}) follows the same steps 
as in the case C-III-c. However, in this case there is no freedom to apply an overall
phase rotation to transform the relative phase of $w_1$ and $w_2$ into two
symmetric phases, since this would make $w_S$ complex. It turns out that 
this vacuum is more constrained than case C-III-c, requiring four relations among
the coefficients of the potential to be obeyed. As a result, the SO(2) rotation 
transforming it into $(be^{i\gamma_1},be^{i\gamma_2},\hat w_S)$ 
automatically leads to $\gamma_1+\gamma_2=0$. Once again 
building the matrix $U$ requires special insight. The necessary SO(2) rotation will
be a function of the $\hat w_i$ and $\sigma_i$, for $i= 1, 2$ and is 
similar to the one of case C-III-c  \cite{Emmanuel-Costa:2016vej} being given 
by $\tan 2 \theta = (\hat w_1^2-\hat w_2^2)/(2\hat w_1\hat w_2 \cos(\sigma_1-\sigma_2))$. 
The alternative procedure of going directly to the Higgs 
basis is also, in this case, the easiest and most 
straightforward way of checking for CP conservation where now a possible 
rotation is:
\begin{equation}
\left( \begin{array}{c}
h_1^\prime \\
h_2^\prime \\
h_S^\prime \\
\end{array}  \right) = \left( \begin{array}{ccc} 
\frac{1}{N_1} (\hat {w_1}  & \hat {w_2} & \hat {w_S})  \\
\frac{1}{N_2} (\hat {w_2} & - \hat {w_1}& 0) \\ 
\frac{1}{N_3} ( \hat {w_1}  &  \hat {w_2} & X) \\
\end{array} \right) \left( \begin{array}{ccc} 
e^{-i\sigma_1} & 0 & 0 \\
0 & e^{-i\sigma_2}  & 0 \\
0 & 0 & 1 
\end{array} \right) 
 \left( \begin{array}{c}
h_1 \\
h_2 \\
h_S \\
\end{array}  \right)
\label{basis3}
\end{equation}
where $1/ N_i$ are normalisation factors and the $X$ is chosen in 
such a way that rows 1 and 3 are also orthogonal. With this transformation only 
$h_1^\prime$ acquires a non-zero vev and the coefficients of the potential 
can all be made real using the freedom to rephase the fields with zero vevs.\\

\noindent
Examples C-III-c and C-IV-e show that searching for a matrix $U$ 
satisfying the constraint of Eq.~(\ref{geo}) may not always be the easiest 
path to check for CP conservation. In particular, as the complexity grows,
it may be more convenient to inspect the potential directly by going 
to the Higgs basis.  \\

\noindent
{\it  The T. D. Lee Model} \\
So far we have shown how to use the Higgs basis to prove that CP is not 
spontaneously broken. In T. D. Lee's two-Higgs-doublet model  \cite{Lee:1973iz}
the potential has the most general form with real coefficients:
\begin{align}
V(\phi)
&=-\lambda_1\phi_1^\dagger\phi_1-\lambda_2\phi_2^\dagger\phi_2 \nonumber \\
&+A(\phi_1^\dagger\phi_1)^2+B(\phi_2^\dagger\phi_2)^2+C(\phi_1^\dagger\phi_1)(\phi_2^\dagger\phi_2) 
+\bar C (\phi_1^\dagger\phi_2)(\phi_2^\dagger\phi_1) 
\nonumber \\
&+\frac{1}{2}[
(\phi_1^\dagger\phi_2)(D\phi_1^\dagger\phi_2+E\phi_1^\dagger\phi_1+F\phi_2^\dagger\phi_2) +\text{h.c.}].
\end{align}
CP is violated spontaneously by vevs of the form $(\rho_1e^{i \theta}, \rho_2)$, in the region of parameters of the potential where $\rho_1$ and $\rho_2$ are 
different from zero and $e^{i \theta} \neq 1$. 
The transformation to the Higgs basis is given by
\begin{equation}
\left( \begin{array}{c} 
\phi_1^\prime \\
\phi_2^\prime \\
\end{array}  \right) = \frac{1}{v} \left( \begin{array}{cc}
1 & 0  \\
0 & e^{i \chi} \\
\end{array} \right) 
\left( \begin{array}{cc} 
\rho_1 & \rho_2 \\
- \rho_2 & \rho_1 \\ 
\end{array} \right) \left( \begin{array}{cc} 
e^{- i \theta} & 0  \\
0 & 1 \\
\end{array} \right) 
 \left( \begin{array}{c}
\phi_1 \\
\phi_2 \\
\end{array}  \right)
\label{basis}
\end{equation}
with $v^2 = \rho_1^2+\rho_2^2$. The potential acquires a new form under this transformation.
Even if the model is initially defined with a bilinear part that is real, this change of basis will generate bilinear terms where the two doublets mix. The coefficients of the bilinear terms $\phi_1^\dagger\phi_2$ and $\phi_2^\dagger\phi_1$ are found to be real only if $\lambda_1=\lambda_2$ or $\sin\chi=0$ or $\sin 2\theta=0$ after performing the change of basis given by eq.~(\ref{basis}).
In each case requiring also the quartic part 
of the potential to be real leads to special conditions on the parameters 
of the potential and therefore does not hold in general.

In the general two-Higgs-doublet model CP can also be violated 
explicitly. This requires the potential to have complex coefficients. 
However, the presence of complex coefficients does not always signal 
CP violation. Again in this case, the 
Higgs basis transformation can be used to check for CP conservation
\cite{Ginzburg:2016tlh}.

\noindent
{\it Concluding remarks:} Both methods presented  in this work
are powerful and rigorous. 
In both cases  there is the need to determine the vevs of the set of Higgs 
doublets. The prescription of Ref.~\cite{Branco:1983tn} is especially 
convenient when working in a basis where there is a symmetry such that the 
matrix $U$ can be easily identified. If that is the case this procedure
is the most direct one of the two. However, this may not always be the case, 
especially  with an increasing number of Higgs doublets. Then the procedure 
described in this paper, which can be  applied in a straightforward manner, 
is a more adequate alternative method to check whether or not CP 
is spontaneously violated. In cases where
spontaneous CP violation is confirmed, the use of weak-basis 
CP-odd invariants that 
signal CP violation, involving vacuum expectation values, can 
provide insight into physical processes that see such effects. 
This type of invariants was used in the case of two Higgs doublets
to study Feynman rules and CP violation \cite{Branco:1999fs}
and also to help distinguish experimentally explicit from spontaneous CP 
violation in the case of two Higgs doublets, with no reference to Yukawa 
couplings \cite{Grzadkowski:2013rza, Grzadkowski:2016szj}. 

It should be pointed out that he usefulness of the transformation to the 
Higgs basis goes beyond the study of the possibility of having 
spontaneous CP violation. If fact, the method proposed in this paper
can also be applied to potentials with complex coefficients.
As pointed out in the introduction couplings that are complex in one basis
may become real in another, and vice versa, and CP may still be
a good symmetry. Once the vevs are known, the same procedure applies to check 
for CP conservation. If CP happens to be violated, in this case, its origin can 
either be explicit or spontaneous.\\

\noindent
{\it Acknowledgments}
The authors thank G. C. Branco and H. Haber for discussions.
PO and MNR thank the CERN Theory Division, where part of this work was done,
for hospitality and partial support.
The authors also thank the University of Bergen and CFTP/IST/University of Lisbon, 
where collaboration visits took place. MNR visited the University of Bergen in 2016 supported by COST Action CA15108 under a Short Term Scientific Mission Grant.
PO is supported by the Research Council of Norway. This work was partially supported by Funda\c c\~ ao para a Ci\^ encia e a Tecnologia (FCT, Portugal) through the projects CERN/FIS-NUC/0010/2015, CFTP-FCT Unit 777 (UID/FIS/00777/2013) which are partially funded through POCTI (FEDER), COMPETE, QREN and EU.

\end{document}